\documentclass[]{elsart}

\usepackage{graphics}
\usepackage{amssymb}

\makeatletter

\providecommand{\LyX}{L\kern-.1667em\lower.25em\hbox{Y}\kern-.125emX\@}
\let\SF@@footnote\footnote
\def\footnote{\ifx\protect\@typeset@protect
    \expandafter\SF@@footnote
  \else
    \expandafter\SF@gobble@opt
  \fi
}
\expandafter\def\csname SF@gobble@opt \endcsname{\@ifnextchar[%]
  \SF@gobble@twobracket
  \@gobble
}
\edef\SF@gobble@opt{\noexpand\protect
  \expandafter\noexpand\csname SF@gobble@opt \endcsname}
\def\SF@gobble@twobracket[#1]#2{}

\usepackage[T1]{fontenc}
\usepackage[latin1]{inputenc}
\usepackage{graphics}

\makeatletter

\usepackage{natbib}
\usepackage{graphicx}
\usepackage{bm}
\usepackage{amsfonts}
\usepackage{amssymb}
\usepackage{amsmath}

\makeatother
\makeatother

\journal{Physica A}

\begin{document}

\begin{frontmatter}

\title{Thermodynamics of fiber bundles}

\author[rennes]{Steven R. Pride\corauthref{cor1}} \ead{Steve.Pride@univ-rennes1.fr} ,
\author[oslo]{Renaud Toussaint}\ead{Renaud.Toussaint@fys.uio.no} 
\address[rennes]{G\'{e}osciences Rennes, Universit\'{e} de Rennes 1, 35042 Rennes Cedex, France}
\address[oslo]{Dept. of Physics, University of Oslo, P.O.Box 1048 Blindern, 0316 Oslo, Norway} 
\corauth[cor1]{Corresponding author. Fax: + 33 2 23 23 60 90}
%\author{Steven R. Pride \corauthref{cor1}} \ead{Steve.Pride@univ-rennes1.fr} 
%\corauth[cor1]{Steven R. Pride}
%\address{G\'{e}osciences Rennes, Universit\'{e} de Rennes 1, 35042 Rennes Cedex, France}

%\author{Renaud Toussaint}\ead{Renaud.Toussaint@fys.uio.no} 
%\address{Dept. of  Physics, University of Oslo, P.O.Box 1048 Blindern, 0316 Oslo, Norway} 

%\date{\today}

\begin{abstract}

A recent theory that determines the properties of disordered
solids as the solid accumulates damage is applied to the special case of fiber
bundles with global load sharing and is shown to be exact in this case. 
The theory postulates that the probability of observing a given emergent
 damage state  is obtained by maximizing the emergent
entropy as defined by Shannon subject to energetic constraints. This theory
yields  the known exact results for the fiber-bundle model with global load 
sharing and   holds for any quenched-disorder distribution. 
It further defines
how the entropy evolves as a function of stress, and shows definitively how the concepts
of temperature and entropy emerge in a problem where all statistics derive 
from the initial quenched disorder.    
 A previously unnoticed phase transition is shown to exist  as the  
entropy goes through a maximum.  In general, this entropy-maximum 
transition  
occurs at a different point in strain history  
than  the 
stress-maximum transition with the precise location depending entirely 
on the quenched-disorder distribution. 
%The  principle manifestation of 
%this transition is that the 
%root-mean-square fluctuation in the number of broken fibers goes through 
%a maximum at this point. 

\end{abstract}

\end{frontmatter}

\section{INTRODUCTION}

The fiber-bundle model with global load sharing is a simple model for failure
in tension introduced almost 80 years ago {[}1--3{]} \nocite{Pei26} \nocite{Dan45} \nocite{Col58} and having  received considerable 
attention and extensions over the past 15 years {[}4--19{]} \nocite{ASL97} \nocite{Dan89} \nocite{HHe94} \nocite{HHa92} \nocite{HKH01} \nocite{HKH02} \nocite{KHH97} \nocite{KZH00} \nocite{KSi82} \nocite{MSB+01} \nocite{MMG+01} \nocite{MGP99} \nocite{MGP00} \nocite{RDP99} \nocite{Rou00} \nocite{Sor89a}.
Although this model may  have little pertinence to  real fibrous systems such
as a rope breaking in tension, it is of interest because it possesses
exact analytical properties. 

In recent work {[}20--22{]} \nocite{TPr02A} \nocite{TPr02B} \nocite{TPr02C}, we have developed a general statistical theory
for determining the properties of a disordered solid that is accumulating irreversible
damage due to cracking under stress. The ensembles in this theory are created  
by considering different realizations of the quenched disorder in the local
breaking strength of the material. Such realizations are made either for the
system as a whole,  
 or of more pertinence to
real systems, by dividing a given system into smaller ``mesovolumes'' and
letting each mesovolume correspond to a different realization of the quenched
disorder. The ensembles  so obtained have
nothing to do with thermal fluctuations (molecular dynamics). Given a system
with  such quenched disorder, our theory determines
the probability of emergent crack states by maximizing Shannon's measure of
disorder subject to  constraints coming from the energetics of
the fracture process. 

In the present paper, we apply this theory to the specific problem of fiber bundles
with global load sharing and demonstrate that for any quenched-disorder distribution, 
it yields the  known
exact results.  
Furthermore, a previously unnoticed phase transition is demonstrated to 
exist where entropy 
goes through a maximum.  This phase transition is distinct from the 
well-known stress maximum 
transition and was noticed in the present  theory because of the 
prominent role played by entropy.  

However, the importance of our theory is not that it yields a new result 
in this old model, but that  
 it applies and yields analytical results about
phase transitions for any  quasi-static damage model;
albeit, for  models involving crack interactions, approximate treatment 
of the functional integrations
may be  required to obtain analytical results (such as renormalization or mean-field
approximations). 
Using our theory, we recently
treated the problem of how the mechanical properties of rocks change due to
cracks arriving and interacting in compressive shear {[}20--22{]}. 
We analytically demonstrated
that the localization transition observed in experiments is a second-order phase
transition when it occurs.

\section{THERMODYNAMICS}

Our theory was originally developed for a disordered solid under the influence
of a stress tensor. For fiber bundles, a much simpler scalar
theory applies and so in this section, the formalism is rederived in this simplified
context. 

\begin{figure}
{\par\centering \resizebox*{6cm}{!}{\includegraphics{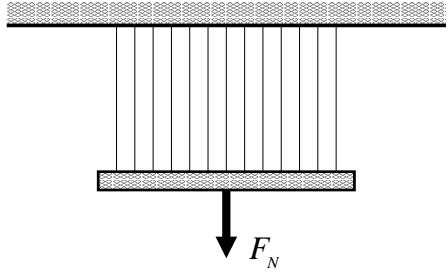} }
\caption{\footnotesize A bundle of \protect\( N\protect \) fibers stretched between
two rigid supports with a load \protect\( F_N\protect \) applied through
 the free support. }
\label{fiberbundle}
\par}
\end{figure}

A fiber bundle is depicted in Fig.\ \ref{fiberbundle}.  A collection of $N$ fibers
are stretched between two rigid supports.  One support is held fixed,
while the other is free to displace.  A load $F_N$ is applied to the bundle
through the free support so that the fibers are in a state of tension.
In this paper,
the load $F_N$ will always be normalized by $N$ to define the overall tension
$\tau = F_N/N$.  
Each non-broken fiber in a bundle has the same length $L$.
If, when $\tau=0$, this length is $L_o$, then the measure of strain
 is $\varepsilon=L/L_o-1$.
Experiments may be performed on the bundle either by controlling 
$\tau$ or $\varepsilon$.

Each of the fibers has the same
Young's modulus which is taken to be unity  so that the
axial strain $\varepsilon$ of each fiber is identical to  the
tension in the fiber.   The $N$ fibers have  strengths
$\varepsilon_1$, $\varepsilon_2$,...$\varepsilon_N$ 
which are independent random variables sampled from a distribution 
\( p(\varepsilon ) \),
whose cumulative distribution is defined
 \( P(\varepsilon )=\int ^{\varepsilon }_{0}p(x)dx \). 
As the strain $\varepsilon$ of the bundle  
 is increased, the individual fibers will break
once their tension (strain) gets to their fixed strength threshold.
The assumption of ``global load sharing'' is that when one of
the fibers breaks at fixed load, all the other fibers will extend by the same
amount thus increasing the tension in each of the surviving fibers
so that the load as a whole is always supported entirely by the surviving
fibers.  
All of this defines the fiber bundle model with global load sharing.
Of interest are the mechanical properties of such bundles as
averaged over all possible realizations of the fiber strengths.

We first need to know the probability \( p_{j} \) of observing one of the realizations
to be in a particular state \( j \) of damage when the ensemble as a whole
is at an applied strain \( \varepsilon  \). In the fiber bundle model, a damage
state \( j \) is defined by which of the \( N \) fibers are broken. One could
define \( j \) using a local order parameter that is 1 if a fiber is intact
and 0 if the fiber is broken. 

Our theory postulates that the fraction \( p_{j} \) of all realizations observed
to be in  state \( j \)  is obtained
by maximizing Shannon's measure of disorder 
\begin{equation}
\label{shannon}
S=-\sum _{j}p_{j}\ln p_{j}
\end{equation}
 subject to constraints. Such constraints must involve the independent variables
of \( S \).   To identify the independent variables, we consider
how both \( S \) and the average energy in the ensemble of bundles change  
as  the strain is increased. 

When  
 \( \varepsilon  \) increases to \( \varepsilon +d\varepsilon  \),  
 there is both a work carried
out in reversibly stretching the fibers and an additional work carried out due
to irreversible fiber breaks. Due to breaking, some of the members of the ensemble
(individual realizations of the disorder) will be led out of their current damage
state and into state \( j \), while others that were in state \( j \) will
transfer to still different states. If there is a difference in the number of
members entering and leaving state \( j \), there
will be a change \( dp_{j} \) in the occupation probability of state \( j \)
and such changes are what cause Shannon's disorder measure \( S \) to change. 

The average energy density (average energy normalized by $N$) in the ensemble 
is given by \( U=\sum _{j}p_{j}E_{j} \). Here, \( E_{j} \) is the energy 
density required
to create state \( j \) at imposed strain  
\( \varepsilon  \) and as averaged over all members that have been led to state
\( j \).   Depending on the breaking strengths of a given realization,
the work performed in arriving at state \( j \) can be different. It is 
through $E_j$ that all dependence on the quenched-disorder distribution 
enters the problem.   The change that
occurs when \( \varepsilon  \) increases to \( \varepsilon +d\varepsilon  \)
is 
\begin{equation}
\label{dU,divided}
dU=\sum _{j}E_{j}dp_{j}+\sum _{j}p_{j}dE_{j}.
\end{equation}
 The first term is the energy expended in changing the disorder over the collection
of realizations. It is thus proportional to the disorder change and can be written
\begin{equation}
\label{dE,advected}
TdS=\sum _{j}E_{j}dp_{j}.
\end{equation}
 The second term is written 
\begin{equation}
\label{dE,incompressible}
f\, d\varepsilon =\sum _{j}p_{j}dE_{j}.
\end{equation}
 and represents both the reversible stretching energy in those members that
did not experience breaks during the deformation increment, as well as the irreversible
energy changes due to all the breaks that did not result in a net change in
the occupation numbers of each state. 

This decomposition of the energy increment can  be thought of  as follows.  
When breaks occur throughout the ensemble
of realizations as \( \varepsilon  \) increases to \( \varepsilon +d\varepsilon  \),
there is a flow of members between the states. This flow involves energy changes
due to fibers breaking in the interval $d\varepsilon$. 
It may be resolved into a uniform ``incompressible''
part whose associated energy is included within \( f\, d\varepsilon  \) as well
as a non-uniform ``compressible'' part associated with more members arriving
in a given state than are leaving that state. The energy associated with the
non-uniform flow between states is entirely contained in \( TdS \).

From these expressions it may be concluded that if \( U \) is to be treated
as a fundamental function, then \( U=U(S,\varepsilon ) \), or equivalently
if \( S \) is to be treated as the fundamental function then \( S=S(U,\varepsilon ) \).
In other words, the independent variables that must be involved in the constraints
on the maximization of \( S \) are \( U \) and \( \varepsilon  \). The proportionality
constants \( T \) and \( f \) are defined 
\begin{equation}
T  =  \left( \frac{\partial U}{\partial S}\right) _{\varepsilon } \, \, 
\mbox{and} \, \, 
f  =  \left( \frac{\partial U}{\partial \varepsilon }\right) _{S}.\label{def,e} 
\end{equation}
The state function \( f \)
is something different than the overall tension \( \tau  \)
since we also have that \( \tau \, d\varepsilon  = dU\).  Thus, in general,  
$(\tau - f) d\varepsilon = T dS$ so that $f \neq \tau$ due to fibers breaking  
in a  positive increment $d\varepsilon$. 
If strain were to be decreased, fibers do not break and so \( dS=0 \)
and the state function \( f \) would be defined using only the purely elastic
part of the energy changes \( dE_{j} \). Changes in \( f \) in this case are  
equivalent to changes in \( \tau  \). 
Last, since we  have taken   $S$ to be extensive 
(proportional to $N$) while $U$ is a density independent of $N$,  $T$ comes out 
being proportional to $1/N$.  This choice is made so that factors of $N$ 
do not clutter the equations that follow. 

The constraint involving \( \varepsilon  \) is  that each non-broken fiber 
throughout the entire ensemble has the same length  which implies  
\( \varepsilon _{j}=\varepsilon  \).   
The constraint involving \( U \) is 
that \( U=\sum _{j}p_{j}E_{j} \). Carrying out the maximization of \( S \)
subject to these constraints using Lagrangian multipliers 
 gives the probability distribution as  
\begin{equation}
\label{prob}
p_{j}(\beta ,\varepsilon )=\frac{e^{-\beta E_{j}(\varepsilon )}}{Z(\beta ,\varepsilon )}
\end{equation}
 where \( \beta =1/T \) and where the partition function \( Z \) is defined
\begin{equation}
\label{partition,funct}
Z(\beta ,\varepsilon )=\sum _{j}e^{-\beta E_{j}(\varepsilon )}.
\end{equation}
 The free energy \( F=F(\beta ,\varepsilon ) \) in this ensemble
(for this set of constraints) is the Legendre transform of \( U \) with respect
to \( S \); \emph{i.e.,} \( F=U-TS \). Differentiation then gives 
\begin{equation}
\label{dF}
dF=\beta ^{-2}Sd\beta + f\, d\varepsilon .
\end{equation}
 Upon introducing \( U=\sum _{j}p_{j}E_{j} \) and \( S=-\sum _{j}p_{j}\ln p_{j} \)
into the Legendre transform \( F=U-S/\beta  \) and using that \( \ln p_{j}=-\beta E_{j}-\ln Z \)
we obtain the standard result 
\begin{equation}
\label{lnZ}
\beta F=-\ln Z.
\end{equation}
 Thus, the standard canonical ensemble emerges in this problem where quenched
disorder alone (not molecular fluctuations) is responsible for the existence
of ensembles. 

An important question in such an approach is whether anything precise can be
said about the temperature \( T=1/\beta  \). Indeed, \( \beta  \) can formally
be found as the solution of a differential equation. This differential equation
is obtained from the previously stated but unused fact that \( dU=\tau d\varepsilon  \) which 
can be written 
\begin{equation}
\label{basis,eq,dif,gene}
\tau =\frac{dU}{d\varepsilon }
=\sum _{j}\frac{dp_{j}}{d\varepsilon }E_{j}+\sum _{j}p_{j}\frac{dE_{j}}{d\varepsilon }.
\end{equation}
 Now from Eq.\ (\ref{prob}), we obtain that 
\begin{equation}
\label{dp,j,deps}
\frac{dp_{j}}{d\varepsilon }=
\left[ -\frac{dE_{j}}{d\varepsilon }\beta -E_{j}
\frac{d\beta }{d\varepsilon }-\frac{d\ln Z}{d\varepsilon }\right] p_{j}, 
\end{equation}
 while from Eqs.\ (\ref{dF}) and (\ref{lnZ}) 
\begin{equation}
\frac{d\ln Z}{d\varepsilon }  =  
-F\frac{d\beta }{d\varepsilon }-\beta \frac{dF}{d\varepsilon }
  =  -U\frac{d\beta }{d\varepsilon }-\beta f.\label{dlnZ,deps} 
\end{equation}
 Since each member is at the same \( \varepsilon  \), each member has its own
\( \tau _{j} \), and so \( \tau =\sum _{j}p_{j}\tau _{j} \). We then obtain
the differential equation for \( \beta  \) in the form 
\begin{equation}
\label{tempequation}
a\frac{d\beta }{d\varepsilon }+b\beta +c=0
\end{equation}
 with coefficients given by 
\begin{equation}
a  =  \sum _{j}p_{j}E_{j}(U-E_{j}); \, \, \, 
b  =  \sum _{j}p_{j}E_{j}\!\left( f-\frac{dE_{j}}{d\varepsilon }\right);  \, \, \,   
c  =  \sum _{j}p_{j}\!\left( \frac{dE_{j}}{d\varepsilon }-\tau _{j}\right).\label{def,c} 
\end{equation}
 Since \( p_{j}=p_{j}(\beta ,\varepsilon ) \), this equation is non-linear
and thus difficult to solve. In the present work, we demonstrate that a proposed
function \( \beta =\beta (\varepsilon ) \) exactly satisfies this equation
and thus is the true ``fiber-bundle'' \( \beta  \). 
To make progress, we next obtain \( E_{j} \) and \( \tau _{j} \) for the specific
problem of fiber bundles with global-load sharing.

\section{FIBER BUNDLE MODEL}

\begin{figure}
{\par\centering \resizebox*{8cm}{4cm}{\includegraphics{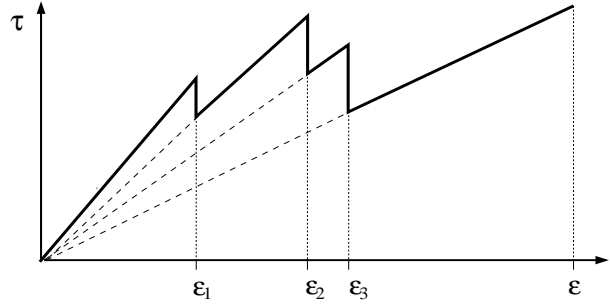}} \par}

\caption{\footnotesize A strain-controlled experiment. 
The solid line is the  load path  followed during the experiment.  
The dashed lines represent the path that would be followed if strain were to 
be decreased at some point during the experiment. When a  
fiber breaks at constant $\varepsilon$, the load $\tau$ must be reduced as 
represented by the vertical drops.  
%When $\varepsilon$ is the control variable, 
%there are no ``avalanches'' as there are when $\tau$ is controlled.  
%But the equivalent of an avalanche is to count how many breaks must occur before 
%the bundle returns to its previous maximum stress level.  Such previous maximums 
%are represented by the horizontal dotted lines.  
% When fiber 1 broke, 
%no avalanche occured.  However, when fiber 2 broke there was a one event avalanche that followed. 
\label{crackhistory}}
\end{figure}

Figure \ref{crackhistory} details the history of how the overall tension $\tau$   
 might evolve when controlled variations in $\varepsilon$  are applied to a 
bundle having   fiber
strengths \( \varepsilon _{1},\varepsilon _{2} \)...\( \varepsilon _{N} \).
In this particular example, \( n=3 \) fibers have broken when 
the strain is at $\varepsilon$.

The load on the bundle is equally shared by the \( N-n \) surviving fibers 
so that 
\begin{equation}
\label{tau_n}
\tau _{n}=\varepsilon \left(1-\frac{n}{N}\right)
\end{equation}
which is a relation  independent of the history; {\em  i.e.} it  depends only on
the actual state of the bundle through the number of broken fibers \( n \), 
 and not on the breaking thresholds 
\( \varepsilon _{1},\varepsilon _{2} \)...\( \varepsilon _{n} \).

The total work density \( E^{p}_{j} \)  for states $j$ consisting 
of $n$ broken fibers is  the area under the charging curve in a particular realization 
(the area under the solid line  
 in Fig.\ \ref{crackhistory})  
\begin{equation}
E^{p}_{j}  =  \int ^{\varepsilon }_{0}\tau (x) \,dx
  =  \sum ^{n}_{m=0}\int ^{\varepsilon _{m+1}}_{\varepsilon _{m}}\tau _{m}(x)\,dx
  =  \sum ^{n}_{m=0}\left( 1-\frac{m}{N}\right) 
\left( \frac{\varepsilon _{m+1}^{2}}{2}-\frac{\varepsilon ^{2}_{m}}{2}\right) 
\nonumber
\end{equation}
where from Eq.\ (\ref{tau_n}) we have used $\tau_m(x)=x(1-m/N)$ and where 
by convention \( \varepsilon _{n+1}=\varepsilon  \) is  the final applied strain.
 A direct recursion  gives exactly  
\begin{equation}
\label{E,j,p}
E_{j}^{p}=\left( 1-\frac{n}{N}\right) \frac{\varepsilon ^{2}}{2}+\sum _{m=1}^{n}
\frac{\varepsilon _{m}^{2}}{2N}.
\end{equation}
 The first term here is the elastic energy that can be reversibly recovered upon 
decreasing the strain while the second term represents the energy  irreversibly  
consumed in the breaking process.  Both contributions  can be directly vizualized in 
Fig.\ \ref{crackhistory}.

Equation (\ref{E,j,p}) is next  averaged over the quenched disorder 
to obtain the average energy  
 \( E_{j} \) needed to create  state $j$. 
Each 
breaking threshold \( \varepsilon _{m} \) is an independant variable, randomly distributed
according to \( p(\varepsilon_m ) \) under the condition 
that \( 0\leq \varepsilon _{m}\leq \varepsilon  \).
They are therefore distributed according to the probability density 
${p(\varepsilon _{m})}/{P(\varepsilon )}$ where 
the normalization factor accounts for the fact that the upper limit  
\( \varepsilon  \) is independent of the threshold values $\varepsilon_m$. 
Thus, averaging  over the quenched disorder gives 
\begin{equation}
\label{def,f,epsilon}
h(\varepsilon) 
=\left\langle \frac{\varepsilon ^{2}_{m}}{2}\right\rangle _{\rm q.d.}
=\frac{1}{P(\varepsilon )}\int ^{\varepsilon }_{0}\frac{x^{2}}{2}p(x)dx
\end{equation}
where  \( h(\varepsilon ) \) designates the average energy that is 
lost when each fiber breaks.  

The average work density (Hamiltonian) required 
 to create the state \( j \), averaged over all realizations
of the quenched disorder is then in general 
\begin{equation}
\label{E,j}
E_{j}=\left( 1-\frac{n_{j}}{N}\right) \frac{\varepsilon ^{2}}{2}+
\frac{n_{j}}{N}h(\varepsilon ) 
= \frac{\varepsilon ^{2}}{2} - \frac{n_{j}}{N} \left[\frac{\varepsilon ^{2}}{2} 
- h(\varepsilon )\right].
\end{equation}
For example, under the special assumption that the breaking strengths are randomly
sampled from a uniform distribution on \( 0\leq \varepsilon _{m}\leq 1 \),  
we have that $ p(\varepsilon )=1$, $ P(\varepsilon )=\varepsilon$,    
$h(\varepsilon )=\varepsilon ^{2}/6 $,  and  
$E_{j}
={\varepsilon ^{2}}/{2}-({n_{j}}/{N}){\varepsilon ^{2}}/{3}$.

\section{Average Properties}

We now apply the above theory and determine the  thermodynamic 
functions   
as a function of  applied strain $\varepsilon$.
The following analysis is valid for any properly normalized 
quenched-disorder distribution $p(\varepsilon)$.  

The cumulative 
distribution $P(\varepsilon) = \int_0^\varepsilon p(x)\, dx$ 
is the probability that any one fiber has broken when the strain is 
at $\varepsilon$.   
Thus, the fraction of all possible realizations having \( n_j \) broken fibers
and  $N-n_j$ unbroken fibers    
is exactly  
\begin{equation}
\label{pexact}
p_{j}^{\textrm{exact}}=(1-P)^{N-n_j}P ^{n_j}.
\end{equation}
This distribution may be written 
\begin{equation}
\label{pexact,reformulated}
p_{j}^{\textrm{exact}}=p_{0}\exp \left[ -n_j\ln \left( \frac{1-P }{P }\right) \right] 
\end{equation}
 where \( p_{0}=(1-P )^{N} \) is the probability of the unbroken
state \( j=0 \). 
Upon using the Hamiltonian of Eq.\ (\ref{E,j}), 
 the postulate of entropy maximization
 predicts this same distribution to be given by  
\begin{equation}
\label{eq:p,j,th,gene}
p^{}_{j}=
p_0 \exp\left[ n_j \frac{\beta}{N} \left(\frac{\varepsilon ^{2}}{2}
- h\right) \right]
\end{equation}
where $p_0=\exp(- \beta \varepsilon^2/2)/Z$ is the probability of 
the unbroken state.

These two distributions are both Boltzmannians in the number $n_j$ 
of broken fibers and are identical  if  
 the temperature $T=1/\beta$ is given by  
\begin{equation}
\label{eq:cond,beta,general}
\beta (\varepsilon )=\frac{-N} 
{\varepsilon ^{2}/2- h(\varepsilon)}
\ln \left
(\frac{1-P(\varepsilon )}{P(\varepsilon )}\right ).
\end{equation}
If it can be shown that this $\beta$ satisfies 
the differential equation of  Eq.\ (\ref{tempequation}), 
then  our theory is exact when  applied 
to fiber bundles with global load sharing. 

From the definition 
 $P(\varepsilon) h(\varepsilon) = \int_0^\varepsilon p(x) x^2/2  \, dx$, one has  
$h(\varepsilon) <  \varepsilon^2/2$ for any distribution $p(\varepsilon)$.
Thus, $\beta(\varepsilon)$ is  a negative increasing function up to the strain point $\varepsilon = 
\varepsilon_\beta$ where it smoothly goes to zero.  The inflection point  $\varepsilon_\beta$ is 
obtained from the condition that $P(\varepsilon_\beta) = 1/2$ and 
defines 
a previously unnoticed phase transition that will be shown to be distinct from the transition 
at peak stress.   
  When $\varepsilon > \varepsilon_\beta$, 
 $\beta$ becomes  a positive increasing function of $\varepsilon$.    

There are two key average properties upon which all the thermodynamic functions  
depend; namely, the average fraction of broken fibers in each bundle \( \langle n_j/N\rangle  \)
and the average of this fraction squared \( \langle (n_j/N)^{2}\rangle  \). Using
the exact probabilities of Eq.\ (\ref{pexact,reformulated}) {[}which is equivalent to using
the \( \beta  \) of Eq.\ (\ref{eq:cond,beta,general}) in our probability law{]}, one obtains
\begin{equation}
\label{technic,average,n}
\left< \frac{n_j}{N}\right> =\sum _{j}p_{j}\frac{n_{j}}{N}=\sum _{n=0}^{N}c_{n}^{N}\frac{n}{N}
P ^{n}(1-P)^{N-n}
\end{equation}
 where \( c_{n}^{N}=N!/[n!(N-n)!] \) defines the number of ways of selecting
\( n \) objects from a group of \( N \) distinguishable items. The binomial
theorem states that
\begin{equation}
\label{binom}
(x+y)^{N}=\sum _{n=0}^{N}c_{n}^{N}x^{n}y^{N-n}.
\end{equation}
 Upon differentiating this equation with respect to \( x \) and then
multiplying by \( x/N \),   gives that  when \( x=P  \) and \( y=1-P  \)
\begin{equation}
\label{average,n,unif}
\left< \frac{n_j}{N}\right> =P
\end{equation}
which is a known result consistent with the meaning of $P$.  
 Differentiating a second time yields
\begin{equation}
\label{aberage,nsquare,unif}
\left< \left( \frac{n_j}{N}\right) ^{2}\right> =P^{2}+\frac{P (1-P )}{N}.
\end{equation}
 Using these two results, the other averages defining the 
thermodynamic variables are easily read off.

The average stress $\tau(\varepsilon)$ is thus obtained to be     
\begin{equation}
\label{eq:tau,gene}
\tau =\sum _{j}p_{j}\tau _{j}=(1-P)\varepsilon  
\end{equation} 
which is initially  an increasing function of $\varepsilon$ having the  slope  
\begin{equation}
\label{slope}
\frac{d\tau}{d\varepsilon} = 1- P - \varepsilon p.
\end{equation}
This slope goes to zero and the stress has a maximum at any strain 
point $\varepsilon_\tau$ that is a 
solution of  $1-P(\varepsilon_\tau) - \varepsilon_\tau p(\varepsilon_\tau) =0$; 
{\em i.e.,} at the point(s) where  
\begin{equation}
\label{eq:cond,peak,gene}
\varepsilon _{\tau} \, p(\varepsilon _{\tau})= 1- \int_0^{\varepsilon_\tau} 
p(x)\, dx
\end{equation}
admits a solution.   This  is a known exact result {[}4,10{]}. \nocite{KHH97}  \nocite{ASL97}  
In general,  we can expect that 
 $\varepsilon _{\tau}\neq \varepsilon_\beta$. The condition required 
for equality of these strain points is that simultaneously $\varepsilon p(\varepsilon) =1/2$ 
and $\int_0^\varepsilon p(x)\, dx =1/2$ which for monotonic distribution functions 
$p(\varepsilon)$ can only occur with the uniform 
distribution $p(\varepsilon)=1$ in which case $\varepsilon _{\tau}= \varepsilon_\beta =1/2$ 
and the two transitions coincide. 
If $p(\varepsilon)$ is a monotonic increasing function of $\varepsilon$, 
then $\varepsilon_\tau 
< \varepsilon_\beta$ while if it is a decreasing function of $\varepsilon$, then 
$\varepsilon_\tau > \varepsilon_\beta$. For non-monotonic  distributions,
 there can be an arbitrary number of stress maximas  and  
 Eq.\ (\ref{eq:cond,peak,gene}) can have either no solutions or multiple solutions   
{[}10{]} \nocite{KHH97}. The nature of the phase transitions at 
the  distinct strain points $\varepsilon_\beta$ and 
$\varepsilon_\tau$ is discussed in the following section.
 
The average energy  in the ensemble is 
\begin{equation}
U  =  \sum _{j}p_{j}E_{j}
  =  (1-P)\frac{\varepsilon ^{2}}{2}+hP.\label{eq:U,gene}
\end{equation}
Again recalling the definition of $h$ from Eq.\ (\ref{def,f,epsilon})  gives 
\begin{equation}
\frac{dU}{d\varepsilon }  =  (1-P)\varepsilon -p\frac{\varepsilon ^{2}}{2}
+\frac{d(hP)}{d\varepsilon }
  =  \tau \label{eq:basis,diffeq,gene} 
\end{equation}
which is also the equation that gives the differential equation for temperature.  
This is sufficient for demonstrating that the  \( \beta (\varepsilon ) \) of 
Eq.\ (\ref{eq:cond,beta,general}) 
satisfies its differential equation.  Nonetheless, as a consistency test,   
 the coefficients $a$, $b$, and $c$  defined in Eq.\ (\ref{def,c}) 
will  be derived and the differential 
equation will  explicitly be shown  to be satisfied.  

To obtain the state function \( f \), we need first
\begin{equation}
\label{eq:dE,j,depsilon}
\frac{dE_{j}}{d\varepsilon }=\varepsilon 
-\frac{n_{j}}{N}\left [\varepsilon - \frac{p}{P}
\left(
\frac{\varepsilon ^{2}}{2} - h \right) \right].
\end{equation}
From  Eq.\ (\ref{dE,incompressible}) and the lemma of 
Eq.\ (\ref{average,n,unif}) we then have 
\begin{equation}
f  =  \sum_j p_j \frac{dE_{j}}{d\varepsilon }
  =  (1-P)\varepsilon +
p\left( 
\frac{\varepsilon ^{2}}{2}-h \right). \label{e,gene} 
\end{equation}
The   variation of the entropy  is obtained from the energy balance as   
\begin{equation}
\frac{dS}{d\varepsilon }  =  \beta \left (\frac{dU}{d\varepsilon }-f\right )
  =  Np\, \ln\left(\frac{1-P}{P}\right ).\label{dS,deps,gene]} 
\end{equation}
Together with the initial condition \( S(0)=0 \), 
 this is readily integrated
 to give 
\begin{equation}
\label{S,gene}
S=-N\left[ P\ln P+(1-P)\ln (1-P)\right].
\end{equation}
This expression is the classical Shannon result for a set of \( N \) random variables
in  two possible states having  probabilities \( P \)
and \( 1-P  \), which is precisely 
the case of the fiber bundle with  global
sharing. This is another consistency check.  
This entropy $S$  increases from zero [total certainty that every member is intact]  
and goes
smoothly through a maximum [total uncertainty as to what state a member may be in] 
at the same strain point \( P(\varepsilon_\beta )=1/2 \) where 
$\beta$  goes to zero.    After the smooth maximum, $S$   
decreases to reach zero if \( P(\varepsilon ) \) reaches \( 1 \) in which case there is 
total certainty that each member is entirely broken.  
We note that Eq.\ (\ref{S,gene}) can also be directly obtained from 
the Shannon formula upon applying the binomial theorem. 

Finally, the free energy $F$ is again obtained from its Legendre transform 
definition $F=U-S/\beta$ to be 
\begin{equation}
F=\frac{\varepsilon^2}{2} - \left( \frac{\varepsilon^2}{2} - h\right) 
\frac{\ln(1-P)}{\ln(1-P) - \ln P}. 
\end{equation}
At the point $\varepsilon_\beta$ defined by $P(\varepsilon_\beta)=1/2$, 
the free energy diverges due to the fact that $\beta(\varepsilon_{\beta}) =0$ 
while $S(\varepsilon_{\beta})$ remains finite.  
So long as $\varepsilon_\beta \neq \varepsilon_\tau$, 
the free energy does not diverge when (if)  $\varepsilon=\varepsilon_\tau$.
In passing, we also note that 
\begin{equation}
\label{Zresult}
Z=\left[(1-P)^{h} P^{-\varepsilon^2/2}\right]^{N/(\varepsilon^2/2 - h)}
\end{equation}
is the exact expression of the partition function.

With the above results established, we now obtain the 
 coefficients \( a,b,c \) of the differential equation for the temperature as 
\begin{eqnarray}
a & = & -\frac{1-P}{N}\!\left (P h^{2}-\frac{\varepsilon ^{2}}{2}\right )^{2}; \,\,\,
b  =  \frac{1-P}{N}
\!\left (Ph^{2}-\frac{\varepsilon ^{2}}{2}\right )\! 
\!\left (P\varepsilon -p\frac{\varepsilon ^{2}}{2}+ph\right ); \nonumber \\
c & = & -p\left ({h} -  
\frac{\varepsilon ^{2}}{2}\right ).\label{c,gene} 
\end{eqnarray}
 Using these together with  Eq.\ (\ref{eq:cond,beta,general})  for \( \beta  \)
 and its derivative
\begin{equation}
\frac{d\beta }{d\varepsilon } = 
 -\frac{N p}{P (1-P) ( h- \varepsilon^2/2)} 
+ 
 N\ln \!\left(\frac{1-P}{P}\right)\! \!
\left[
\frac{\varepsilon + (h- \varepsilon^2/2)p/P}
{(h-\varepsilon^2/2)^2}
\right] 
\label{dbeta,deps,gene} 
\end{equation}
shows that the differential equation $a \,d\beta/d\varepsilon + b\beta +c=0$ 
is exactly satisfied. 

\section{Phase Transitions}
\subsection{The entropy maximum transition}
The strain  point $\varepsilon_\beta$  defined by $P(\varepsilon_\beta) =1/2$ 
is where  simultaneously $\beta=0$,   the entropy is a maximum, and the free energy diverges.  
 It is  distinct from the stress-maximum transition(s)  
 at $\varepsilon_\tau$.   The interpretation of  $\varepsilon=\varepsilon_\beta$ 
as a phase transition is natural, since  the most probable configuration of a bundle 
suddenly jumps  from being entirely  intact  to  entirely broken. 
What are the measurable manifestations of 
this transition at $\varepsilon_\beta$?  

Since the entropy is a maximum at this point, the fluctuations between realizations should also 
be a maximum as we now demonstrate.  Define the fraction of broken fibers in state 
$j$ to be   
$\rho _{j}={n_{j}}/{N}$, 
and define the average of this fraction to be  
$\rho = \langle \rho_j \rangle = P$ 
where $P=P(\varepsilon)$ is again the probability of a fiber being broken. 
The quantity of interest here is the root-mean-square   fluctuation  
$\Delta \rho$  in the fraction of broken fibers  given by  
\begin{equation}
\Delta \rho = \sqrt{\left\langle \Delta \rho_j ^{2}\right\rangle}   =  
\sqrt{\left\langle \rho_j ^{2}\right\rangle -\rho ^{2}} =  
\frac{\sqrt{P(1-P)}}{\sqrt{N}}
\end{equation}
which indeed goes through 
a maximum at $P(\varepsilon_\beta) = 1/2$ as  expected.  
  This maximum is something that can be directly  measured 
in numerical experiments on fiber bundles but has never before been commented on.  
The reason it has been  discovered in the present theory 
is   because  entropy  
and temperature are  explicitly present. 
We encourage someone to 
numerically measure $\Delta \rho$ and to verify that 
it is a maximum at the  transition point $\varepsilon_\beta$.
Recall that for monotonic quenched-disorder distributions,
if $p(\varepsilon)$ is a decreasing function (more weak fibers than strong
fibers), then $\varepsilon_\beta < \varepsilon_\tau$.  So the   
 numerical observation in this case  should be that   $\Delta \rho$ goes 
 through a maximum
 prior to peak stress.  Equivalent comments hold
when  $p(\varepsilon)$ is an increasing function and $\varepsilon_\beta > \varepsilon_\tau$.

The other  fluctuations that 
are potentially of interest include the root-mean-square stress fluctuation 
\begin{equation}
\Delta \tau = \sqrt{\left\langle \Delta \tau_j^{2}\right\rangle}   =  
\sqrt{\left\langle \tau ^{2}_{j}\right\rangle -\tau ^{2}}
  =  \varepsilon \Delta \rho
\end{equation}
and the root-mean-square energy fluctuation 
\begin{equation}
\Delta U = \sqrt{\left\langle \Delta E_j^{2}\right\rangle}   =  
\sqrt{\left\langle E^{2}_{j}\right\rangle -U^{2}}  
  =  \left (\frac{\varepsilon ^{2}}{2}-h\right ) \Delta \rho, 
\end{equation}
but since these are simply proportional to $\Delta \rho$ it seems that 
the interesting signature of this phase transition is the maximum in 
$\Delta \rho$.

The ``order'' of this transition is not classically defined 
in the Ehrenfest scheme; however, it seems inappropriate to call 
it  a continuous transition because   
the free energy  is singular at $\varepsilon_\beta$ as shown above. 
Nonetheless, in the limit as $\varepsilon \rightarrow \varepsilon_\beta$ the singular 
part of $F$  diverges  according to the scaling law   
\begin{equation}
F_s = -\frac{\ln 2}{8} \frac{[\varepsilon_\beta^2 - 
h(\varepsilon_\beta)/2] }
{p(\varepsilon_\beta)} \, (\varepsilon - \varepsilon_\beta)^{-1} 
\end{equation}
which  has a (trivial) universal exponent. 
Since $F$ and its derivatives are difficult to numerically 
measure, the principle manifestation of this phase transition remains  
 $\Delta \rho$ going through a maximum at a point $\varepsilon_\beta 
\neq \varepsilon_\tau$.

\subsection{The stress maximum transition}
The   phase transition 
at peak stress $\varepsilon = \varepsilon_\tau$ is the one that researchers 
up to now have focused on.    
Since $S$ and its derivatives remain continuous and finite there, an attempt 
to classify this transition as being first-order or second-order 
is meaningless. The  label of ``continuous'' transition seems the 
most appropriate.    

There is   universality at this 
transition due to the fact that for any  monotonic  
 analytic quenched-disorder distribution,  
$\tau(\varepsilon)$ is a smooth analytic function   
 so that  upon developing this function in 
the neighborhood of its  maximum one is guaranteed 
$|\tau - \tau(\varepsilon_\tau)| \sim  
|\varepsilon - \varepsilon_\tau|^2$ with the exponent of course being independent of 
the distribution $p(\varepsilon)$ or other model parameters.  
Such development
of an analytic function  about a critical point is the way any mean-field
theory acquires a universal scaling law.  
However,  Kloster, Hansen and Hemmer 
{[}10{]} \nocite{KHH97} do demonstrate that  a non-analytic quenched-disorder 
distribution can lead to highly non-analytic stress-strain behavior that 
thus falls outside the quadratic  universality class.
Also, in other damage models in
which elastic interaction between cracks (damage points)  is
important, the simple observation of a model exhibiting an averaged  stress maximum does
not by itself guarantee an exactly quadratic  stress-strain relation in the neighborhood
of the maximum.  Such a relation may  be
non-analytic due to a diverging correlation length in the correlation between
cracks.

The quadratic  stress maximum means that $d\varepsilon/d\tau$ diverges as  
$|\tau - \tau(\varepsilon_\tau)|^{-1/2}$ {[}11,19{]} \nocite{Sor89a} \nocite{KZH00}. Accordingly,  
the average rate at which the fraction of broken fibers increases with stress 
 $d\rho/d\tau = p\,  d\varepsilon/d\tau$ also diverges 
as $|\tau - \tau(\varepsilon_\tau)|^{-1/2}$ {[}4,15{]} \nocite{ASL97} \nocite{MGP00}.   
Using additional considerations not developed in this paper, one can  futher show 
that the average size of avalanches also diverges as 
$|\tau - \tau(\varepsilon_\tau)|^{-1/2}$ {[}10,15,17{]}, 
 \nocite{KHH97} \nocite{RDP99} \nocite{MGP00}
 and that the distribution $D(\Delta)$ of the size $\Delta$ 
of the avalanches scales as $D(\Delta) \sim \Delta^{-5/2}$ {[}6,7,10,17{]}  at the stress maximum. 
\nocite{HHe94} \nocite{HHa92} \nocite{KHH97} \nocite{RDP99}  
These are the principal observable characteristics of the stress-maximum transition.

\section{Summary and Conclusions}
Two principal results  have been obtained in this paper.  First, it was  demonstrated
 that 
for any quenched-disorder distribution  used in 
 the fiber-bundle model with global load sharing, the 
probability  of the emergent damage states may be exactly 
calculated by maximizing Shannon's entropy subject to constraints.  
It is through the constraints  
that the nature  of the quenched-disorder distribution 
enters the macroscopic thermodynamics.      

Second, a previously unnoticed phase transition occurs in the 
fiber-bundle model with global load sharing when 
the entropy goes through a maximum.  This phase transition is 
distinct from the stress-maximum transition; although,  for the uniform 
quenched-disorder distribution,   the two transitions coincide.
The principle manifestation of this transition is that the 
root-mean-square fluctuation in the number of broken fibers 
will go through a maximum, which is a quantity that can be 
directly measured in numerical experiments. 

To conclude, 
we postulate that the probability of emergent states can always 
be calculated through entropy maximization for any damage model of interest  
including models in which there is elastic interaction between the local 
damage (order) parameters.  Damage  models with order-parameter interactions 
are not normally considered 
amenable to analytical treament;  however,  using our approach a rather 
standard canonical-ensemble partition function emerges and 
the various functional integration procedures  available for 
studying the partition function in the neighborhood of critical points 
may be employed.   
  
Such generality is the principal utility of our approach.  
A concern is the ability to produce an expression for the model temperature. 
The   temperature   can be found in principle as               
the solution of a well-posed initial-value problem.  Unfortunately, 
the differential problem is highly non-linear and thus 
difficult to solve.
Because stress maxima and entropy maxima do not normally coincide,  
 a   stress-maximum transition may be studied  
by  simply assuming  the temperature to be well behaved in the neighborhood of  
 the stress maxima.   
But if an explicit expression for the temperature is desired, 
the following approach can be employed. 

In  models involving 
crack interactions, there are always a subset of dilute states in which 
the interactions are negligible.  The exact probabilities of such states 
can usually be determined from the quenched-disorder distribution alone. 
A comparison 
between the probabilities calculated from entropy maximization and 
such exact probabilities then determines the model temperature 
to be used in the Boltzmannian. 
  With the temperature so defined, 
and with a proper model Hamiltonian in hand, 
the partition function can be analyzed.


\begin{thebibliography}{22} 

\expandafter\ifx\csname natexlab\endcsname\relax\def\natexlab#1{#1}\fi
\expandafter\ifx\csname bibnamefont\endcsname\relax
  \def\bibnamefont#1{#1}\fi
\expandafter\ifx\csname bibfnamefont\endcsname\relax
  \def\bibfnamefont#1{#1}\fi
\expandafter\ifx\csname url\endcsname\relax
  \def\url#1{\texttt{#1}}\fi
\expandafter\ifx\csname urlprefix\endcsname\relax\def\urlprefix{URL }\fi
\providecommand{\bibinfo}[2]{#2}
\providecommand{\eprint}[2][]{\url{#2}}

\bibitem{Pei26}
\bibinfo{author}{\bibfnamefont{F.T.}~\bibnamefont{Peirce}},
  \bibinfo{journal}{J. Text. Ind.} \textbf{\bibinfo{volume}{17}},
  \bibinfo{pages}{355} (\bibinfo{year}{1926}).

\bibitem{Dan45}
\bibinfo{author}{\bibfnamefont{H.E.}~\bibnamefont{Daniels}},
  \bibinfo{journal}{Proc. R. Soc. A} \textbf{\bibinfo{volume}{183}},
  \bibinfo{pages}{404} (\bibinfo{year}{1945}).

\bibitem{Col58}
\bibinfo{author}{\bibfnamefont{B.D.}~\bibnamefont{Coleman}},
  \bibinfo{journal}{J. Appl. Phys.} \textbf{\bibinfo{volume}{29}},
  \bibinfo{pages}{968} (\bibinfo{year}{1958}).

\bibitem{ASL97}
  \bibinfo{author}{\bibfnamefont{J.V.}~\bibnamefont{Andersen}}, 
 {\bibfnamefont{D.}~\bibnamefont{Sornette}}
 \bibnamefont{and}
  \bibinfo{author}{\bibfnamefont{K.W.}~\bibnamefont{Leung}},
  \bibinfo{journal}{Phys. Rev. Lett.} \textbf{\bibinfo{volume}{78}},
  \bibinfo{pages}{2140} (\bibinfo{year}{1997}).

\bibitem{Dan89}
\bibinfo{author}{\bibfnamefont{H.E.}~\bibnamefont{Daniels}},
  \bibinfo{journal}{Adv. Appl. Prob.} \textbf{\bibinfo{volume}{21}},
  \bibinfo{pages}{315} (\bibinfo{year}{1989}).

\bibitem{HHe94}
\bibinfo{author}{\bibfnamefont{A.}~\bibnamefont{Hansen}} \bibnamefont{and}
  \bibinfo{author}{\bibfnamefont{P.C.}~\bibnamefont{Hemmer}},
  \bibinfo{journal}{Phys. Lett. A} \textbf{\bibinfo{volume}{184}},
  \bibinfo{pages}{394} (\bibinfo{year}{1994}).

%\bibitem{HPh78}
%\bibinfo{author}{\bibfnamefont{D.G}~\bibnamefont{Harlow}} \bibnamefont{and}
%\bibinfo{author}{\bibfnamefont{S.L.}~\bibnamefont{Phoenix}},
%\bibinfo{journal}{J. Composite Mater.} \textbf{\bibinfo{volume}{12}},
%\bibinfo{pages}{195} (\bibinfo{year}{1978}).

\bibitem{HHa92}
\bibinfo{author}{\bibfnamefont{P.C.}~\bibnamefont{Hemmer}} \bibnamefont{and}
  \bibinfo{author}{\bibfnamefont{A.}~\bibnamefont{Hansen}},
  \bibinfo{journal}{J. Appl. Mech.} \textbf{\bibinfo{volume}{59}},
  \bibinfo{pages}{909} (\bibinfo{year}{1992}).

\bibitem{HKH02}
\bibinfo{author}{\bibfnamefont{R.C.}~\bibnamefont{Hidalgo}},
  \bibinfo{author}{\bibfnamefont{F.}~\bibnamefont{Kun}}, \bibnamefont{and}
  \bibinfo{author}{\bibfnamefont{H.J.}~\bibnamefont{Herrmann}},
  \bibinfo{journal}{Phys. Rev. E} \textbf{\bibinfo{volume}{65}},
  \bibinfo{pages}{032502} (\bibinfo{year}{2002}).

\bibitem{HKH01}
\bibinfo{author}{\bibfnamefont{R.C.}~\bibnamefont{Hidalgo}},
  \bibinfo{author}{\bibfnamefont{F.}~\bibnamefont{Kun}}, \bibnamefont{and}
  \bibinfo{author}{\bibfnamefont{H.J.}~\bibnamefont{Herrmann}},
  \bibinfo{journal}{Phys. Rev. E} \textbf{\bibinfo{volume}{64}},
  \bibinfo{pages}{066122} (\bibinfo{year}{2001}).

\bibitem{KHH97}
\bibinfo{author}{\bibfnamefont{M.}~\bibnamefont{Kloster}},
  \bibinfo{author}{\bibfnamefont{A.}~\bibnamefont{Hansen}}, \bibnamefont{and}
  \bibinfo{author}{\bibfnamefont{P.C.}~\bibnamefont{Hemmer}},
  \bibinfo{journal}{Phys. Rev. E} \textbf{\bibinfo{volume}{56}},
  \bibinfo{pages}{2615} (\bibinfo{year}{1997}).

\bibitem{KZH00}
\bibinfo{author}{\bibfnamefont{F.}~\bibnamefont{Kun}},
  \bibinfo{author}{\bibfnamefont{S.}~\bibnamefont{Zapperi}}, \bibnamefont{and}
  \bibinfo{author}{\bibfnamefont{H.J.}~\bibnamefont{Herrmann}},
  \bibinfo{journal}{Eur. Phys. J. B} \textbf{\bibinfo{volume}{17}},
  \bibinfo{pages}{269} (\bibinfo{year}{2000}).


\bibitem{KSi82}
\bibinfo{author}{\bibfnamefont{D.}~\bibnamefont{Krajcinovic}}, \bibnamefont{and}
  \bibinfo{author}{\bibfnamefont{M.A.G.}~\bibnamefont{Silva}},
  \bibinfo{journal}{Int. J. Solids Structures} \textbf{\bibinfo{volume}{18}},
  \bibinfo{pages}{551} (\bibinfo{year}{1982}).

\bibitem{MSB+01}
\bibinfo{author}{\bibfnamefont{I.L.}~\bibnamefont{Menezes-Sobrinho}},
 \bibinfo{author}{\bibfnamefont{A.T.}~\bibnamefont{Bernardes}},
 \bibnamefont{and}
  \bibinfo{author}{\bibfnamefont{J.G.}~\bibnamefont{Moreira}},
  \bibinfo{journal}{Phys. Rev. E} \textbf{\bibinfo{volume}{63}},
  \bibinfo{pages}{025104(R)} (\bibinfo{year}{2001}).

\bibitem{MMG+01}
  \bibinfo{author}{\bibfnamefont{L.}~\bibnamefont{Moral}}, {\bibfnamefont{Y.}~\bibnamefont{Moreno}}
 \bibnamefont{and}
  \bibinfo{author}{\bibfnamefont{A.F.}~\bibnamefont{Pacheco}},
  \bibinfo{journal}{Phys. Rev. E} \textbf{\bibinfo{volume}{63}},
  \bibinfo{pages}{066106} (\bibinfo{year}{2001}).

\bibitem{MGP00}
  \bibinfo{author}{\bibfnamefont{Y.}~\bibnamefont{Moreno}}, {\bibfnamefont{J.B.}~\bibnamefont{Gomez}}
 \bibnamefont{and}
  \bibinfo{author}{\bibfnamefont{A.F.}~\bibnamefont{Pacheco}},
  \bibinfo{journal}{Phys. Rev. Lett.} \textbf{\bibinfo{volume}{85}},
  \bibinfo{pages}{2865} (\bibinfo{year}{2000}).

\bibitem{MGP99}
  \bibinfo{author}{\bibfnamefont{Y.}~\bibnamefont{Moreno}}, {\bibfnamefont{J.B.}~\bibnamefont{Gomez}}
 \bibnamefont{and}
  \bibinfo{author}{\bibfnamefont{A.F.}~\bibnamefont{Pacheco}},
  \bibinfo{journal}{Physica A} \textbf{\bibinfo{volume}{274}},
  \bibinfo{pages}{400} (\bibinfo{year}{1999}).

\bibitem{RDP99}
  \bibinfo{author}{\bibfnamefont{S.}~\bibnamefont{Roux}}, {\bibfnamefont{A.}~\bibnamefont{Delaplace}}
 \bibnamefont{and}
  \bibinfo{author}{\bibfnamefont{G.}~\bibnamefont{Pijaudier-Cabot}},
  \bibinfo{journal}{Physica A} \textbf{\bibinfo{volume}{270}},
  \bibinfo{pages}{35} (\bibinfo{year}{1999}).

\bibitem{Rou00}
  \bibinfo{author}{\bibfnamefont{S.}~\bibnamefont{Roux}},
  \bibinfo{journal}{Phys. Rev. E} \textbf{\bibinfo{volume}{62}},
  \bibinfo{pages}{6164} (\bibinfo{year}{2000}).

\bibitem{Sor89a}
\bibinfo{author}{\bibfnamefont{D.}~\bibnamefont{Sornette}},
  \bibinfo{journal}{J. Phys. A} \textbf{\bibinfo{volume}{22}},
  \bibinfo{pages}{L243} (\bibinfo{year}{1989}).

\bibitem{TPr02A}
\bibinfo{author}{\bibfnamefont{R.}~\bibnamefont{Toussaint}} \bibnamefont{and}
  \bibinfo{author}{\bibfnamefont{S.R.}~\bibnamefont{Pride}},
  \bibinfo{journal}{Phys. Rev. E}  (\bibinfo{year}{2002}),
  \bibinfo{note}{(in press)}.

\bibitem{TPr02B}
\bibinfo{author}{\bibfnamefont{R.}~\bibnamefont{Toussaint}} \bibnamefont{and}
  \bibinfo{author}{\bibfnamefont{S.R.}~\bibnamefont{Pride}},
  \bibinfo{journal}{Phys. Rev. E}  (\bibinfo{year}{2002}),
  \bibinfo{note}{(in press)}.

\bibitem{TPr02C}
\bibinfo{author}{\bibfnamefont{R.}~\bibnamefont{Toussaint}} \bibnamefont{and}
  \bibinfo{author}{\bibfnamefont{S.R.}~\bibnamefont{Pride}},
  \bibinfo{journal}{Phys. Rev. E}  (\bibinfo{year}{2002}),
  \bibinfo{note}{(in press)}.

\end{thebibliography}
\end{document}